\documentclass[aps,prd,superscriptaddress,amsfonts,amssymb,amsmath,eqsecnum,nofootinbib,twocolumn,floatfix]{revtex4}
 \usepackage{color,graphicx}
 \usepackage[utf8]{inputenc}
  \usepackage[T1]{fontenc}
 \usepackage[english]{babel}
 \definecolor{darkblue}{rgb}{0,0,0.7}
\definecolor{darkred}{rgb}{0.7,0,0}
\definecolor{darkgreen}{rgb}{0,0.4,0}
\usepackage[unicode, colorlinks, citecolor=darkblue, linkcolor=darkred, urlcolor=blue]{hyperref}

\allowdisplaybreaks
\begin{document}

\author{Alexandr Karpenko}
\affiliation{Faculty of Physics, M.V. Lomonosov Moscow State University, Leninskie Gory, Moscow 119991,  Russia}

\author{Sergey P. Vyatchanin}
\affiliation{Faculty of Physics, M.V. Lomonosov Moscow State University, Leninskie Gory, Moscow 119991,  Russia\\
     Quantum Technology Centre, M.V. Lomonosov Moscow State University, Leninskie Gory, Moscow 119991,  Russia}

\date{\today}

\title{Combination of dissipative and dispersive coupling \\  in the cavity optomechanical systems}

\begin{abstract}
     An analysis is given for the Fabry-Perot cavity  having a combination of dissipative and dispersive optomechanical coupling. It is established that the combined coupling leads to optical rigidity.  At the same time, this rigidity appears in systems with the combined coupling on the resonant pump, which is not typical for pure dispersive and dissipative couplings. A proposal is made to use this system to detect small signal forces with better sensitivity than SQL. It is also demonstrated that this optomechanical system can create ponderomotive squeezing with controllable parameters over a wider range than ponderomotive squeezing using dispersive coupling.
\end{abstract}

\maketitle

\section{Introduction}

Optomechanics is studying the fundamental sensitivity limitations
in measuring position of test mass. This sensitivity can be very
high. For example, a relative mechanical displacement detected can
be smaller than the size of proton. This feature is widely used in
gravitational wave detectors
\cite{aLIGO2013,aLIGO2015,MartynovPRD16,AserneseCQG15,
DooleyCQG16,AsoPRD13}, in magnetometers \cite{ForstnerPRL2012,
LiOptica2018}, and in torque sensors \cite{WuPRX2014, KimNC2016,
AhnNT2020}.

The fundamental limitation is provided by the quantum noise. In
conventional scheme of resonantly pumped Fabry-Perot (FP) cavity
with movable end mirror (the test mass) the phase of light,
reflected from the cavity, contains information on position of the
test mass. The limit sensitivity is restricted by well known
standard quantum limit (SQL) \cite{Braginsky68,BrKh92}, which is
an interplay between phase fluctuations of incident light (the
measurement of error) and the Lebedev' fluctuation light pressure
force (back action).

SQL was investigated in many systems ranging from macroscopic
kilometer-size gravitational wave detectors
\cite{02a1KiLeMaThVyPRD} to microcavities
\cite{Kippenberg08,DobrindtPRL2010}. Detecting classical force
acting on a test mass in optomechanical systems is an example of
measurements restricted by SQL. It can be surpassed by applying a
variational measurement \cite{93a1VyMaJETP, 95a1VyZuPLA,
02a1KiLeMaThVyPRD}, or squeezed light input \cite{LigoNatPh11,
LigoNatPhot13, TsePRL19, AsernesePRL19, YapNatPhot20, YuArxiv20,
CripeNat19}, or optomechanical speed measurement
\cite{90BrKhPLA,00a1BrGoKhThPRD}, or optical spring
\cite{99a1BrKhPLA, 01a1KhPLA}. SQL can be also avoided using
coherent quantum noise cancellation \cite{TsangPRL2010,
PolzikAdPh2014, MollerNature2017}.

There are two types of opto-mechanic coupling, namely:  dispersive and dissipative coupling. In dispersive coupling displacing mirror is changing normal cavity frequency, whereas in dissipative coupling displacing test mass brings about a change in the input mirror transmittance altering, thereby, the cavity relaxation rate. Dissipative coupling was  proposed theoretically \cite{ElstePRL2009} and confirmed experimentally \cite{LiPRL2009,WeissNJP2013,WuPRX2014, HryciwOpt2015} about a decade ago. This phenomenon was investigated in numerous optomechanical systems, including the FP interferometer \cite{LiPRL2009,WeissNJP2013,WuPRX2014, HryciwOpt2015}, the Michelson-Sagnac interferometer (MSI) \cite{XuerebPRL2011, TarabrinPRA2013, SawadskyPRL2015, 16a1PRAVyMa, 19a1JPhyBNaVy, 20PRAKaVy}, and ring resonators \cite{HuangPRA2010,HuangPRA2010b}. It was demonstrated that an optomechanical transducer based on dissipative coupling  allows realizing quantum speed meter which, in turn, helps to avoid SQL \cite{16a1PRAVyMa}.

The natural question is to what extent the combination of
dispersive and dissipative coupling can improve the sensitivity of
optomechanical system to detect the signal of displaced test mass.
It is known that squeezing output quadratures  is dramatically
different for purely dispersive and dissipative coupling
\cite{16a1PRAVyMa, 20PRAKaVy}. Seemingly, their combination does
not look promising, but this conclusion is not correct.

In this paper we analyse a FP cavity featuring a combination of
these different types of coupling and demonstrate that SQL can be
surpassed. The physical reason is the optical rigidity formed by
the combination of both dispersive and dissipative coupling. We
also demonstrate that this combined coupling gives possibility to
obtain ponderomotive frequency dependent squeezing with controllable parameters over a
wider range is compared with ponderomotive squeezing using
dispersive coupling. Such frequency depended squeezing can be used in laser gravitational wave antennas.

\section{Model}
We consider 1D FP cavity. Its optical mode with eigenfrequency
$\omega_0$ is pumped using resonant light (the pump frequency
$\omega_p = \omega_0$). The optical mode is coupled with the
mechanical system represented by a free mass m. Eigenfrequency
$\omega$ of cavity and relaxation rate $\gamma$ of the optical
mode depend on test mass displacement $y$.
Signal $F_s$ acts on the free mass changing its position.

The description of dissipation in FP cavity is known (for example,
see \cite{Walls2008}), especially in case of dispersive coupling.
Here we use generalization for combined (dispersive and
dissipative) coupling. The Hamiltonian of this a system can be
expressed as
\begin{subequations}
    \label{Hamilt}
\begin{align}
&\hat{H}=\hbar\omega_0(1+\xi\hat{y})\hat{a}_c^{\dagger}\hat{a}_c+\frac{\hat{p}^2}{2m}+\hat{H}_{\gamma}+\hat{H}_{T}-F_s\hat{y},\\
&\hat{H}_{T}=\int\limits_0^\infty\hbar\omega\, \hat{b}_{\omega}^{\dagger}\hat{b}_{\omega}\frac{d\omega}{2\pi},\\
&\hat{H}_{\gamma}=-i\hbar\sqrt{\gamma}\int\limits_0^\infty\left(\hat{b}_{\omega}\hat{a}_c^{\dagger}-\hat{a}_c\hat{b}_{\omega}^{\dagger}\right)
 \frac{d\omega}{2\pi};\\
&\gamma=\gamma_0\left(1+\eta\hat{y}\right), \quad
 \sqrt{\gamma}\simeq\sqrt{\gamma_0}\left(1+\frac{\eta}{2}\hat{y}\right).
\end{align}
 \end{subequations}
Here $\hat{p}$ is the momentum of the test mass, $\hat{a}_c$
$\hat{a}_c^{\dagger}$ are annihilation and creation operators
describing the intracavity optical field, $\hat{H}_{T}$ is the
Hamiltonian of the electromagnetic field outside the cavity
(thermal bath and pump), $\hat{H}_{\gamma}$ describes coupling
between intracavity and extracavity optical fields, $\xi$ and
$\eta$ are the coefficients of dispersive and dissipative
coupling, respectively.

From the Hamiltonian \eqref{Hamilt} we obtain a set of equations
describing the time evolution of the optomechanical system
\begin{subequations}
    \label{eq1}
    \begin{align}
    &\dot{\hat{a}}_c= -i\omega_0(1+\xi\hat{y})\hat{a}_c-\sqrt{\gamma}\int\limits_0^\infty\hat{b}_{\omega}\frac{d\omega}{2\pi},\\
    &\dot{\hat{b}}_{\omega}=-i\omega\hat{b}_{\omega}+\sqrt{\gamma}\hat{a}_{c},\\
    &\ddot{\hat{y}}=-\frac{\hbar\omega_0\xi}{m}\hat{a}^{\dagger}_{c}\hat{a}_c+\frac{F_s}{m}-i\frac{\sqrt{\gamma_0}\eta\hbar}{2m}\int\limits_0^\infty\left(\hat{b}_{\omega}\hat{a}_c^{\dagger}-\hat{a}_c\hat{b}_{\omega}^{\dagger}\right)\frac{d\omega}{2\pi}
    \nonumber
    \end{align}
\end{subequations}

We present the annihilation operators of the input and intracavity optical field through slow amplitudes as

\begin{align}
\label{eq2}
\hat{a}_c(t)\Rightarrow\hat{a}_c(t)e^{-i\omega_0t}, \quad \hat{b}_{\omega}(t)\Rightarrow\hat{b}_{\omega}(t)e^{-i\omega t}.
\end{align}
We substitute \eqref{eq2} into the system of equations \eqref{eq1} and obtain relations for slow amplitudes and displacement of the probe mass
\begin{subequations}
    \label{eq3}
\begin{align}
&\dot{\hat{a}}_c=-i\omega_0\xi\hat{y}\hat{a}_c+\sqrt{\gamma}\hat{a}_{in}-\frac{\gamma}{2}\hat{a}_c,\\
&\dot{\hat{a}}_c=-i\omega_0\xi\hat{y}\hat{a}_c-\sqrt{\gamma}\hat{a}_{out}+\frac{\gamma}{2}\hat{a}_c,\\
&\ddot{\hat{y}}=-\frac{\hbar\omega_0\xi}{m}\hat{a}^{\dagger}_{c}\hat{a}_c+\frac{F_s}{m}-i\frac{\sqrt{\gamma_0}\eta\hbar}{2m}(\hat{a}_{in}\hat{a}_c^{\dagger}-\hat{a}_c\hat{a}_{in}^{\dagger}),\\
\label{eq4}
&\hat{a}_{in}=-\int\limits_{-\infty}^{\infty}\hat{b}(t_0)e^{-i\Omega t}\frac{d\Omega}{2\pi},\\
\label{eq5} &\hat{a}_{out}=\int\limits_{-\infty}^{\infty}\hat{b}(t_1)e^{-i\Omega t}\frac{d\Omega}{2\pi}.
\end{align}
\end{subequations}
Here $t_0$ and $t_1$ are some initial and final time points
respectively. In these equations we present $\omega$ as a sum
$\omega=\omega_0+\Omega$, where $\omega_0$ is the pump frequency
and $\Omega$ is the spectral frequency of the signal $F_s$ (here
$\hat{b}(t)=\hat{b}_{\omega_0+\Omega}(t)$). It is much smaller
than $\omega_0\sim 10^{-15}Hz$, because we formally extend the
integrals over $\Omega$ to infinity in the equations \eqref{eq4}
and \eqref{eq5} to simplify notation.

We can determine the relation between the input and output fields
from equations \eqref{eq3}:
\begin{align}
\label{eq6}
\hat{a}_{in}+\hat{a}_{out}=\sqrt{\gamma}\hat{a}_c.
\end{align}

Below we express amplitudes as a large constant amplitudes
(denoted by capital letters) plus small amplitudes (denoted by the
same letters in low-case) to describe the noise and signal
components:
\begin{align}
\label{a1}
\hat{a}_{c}=A_0+\hat{a}_0; \quad \hat{a}_{in}=A+\hat{a}; \quad \hat{a}_{out}=A_1+\hat{a}_1.
\end{align}
Here and below we assume that the input wave is in coherent state,
so operator $\hat a$ describes the vacuum fluctuation wave having
the following commutator and correlator
\begin{align}
\label{eq7}
\left[\hat a(t), \hat a^\dag(t')\right] &=   \delta(t-t'),\ \
\left\langle\hat a(t) \hat a^\dag(t')\right\rangle = \delta(t-t')
\end{align}

The Fourier transform can be defined as follows
\begin{align}
\hat a(t) &= \int_{-\infty}^\infty a(\Omega) \, e^{-i\Omega t}\, \frac{d\Omega}{2\pi}
\end{align}
and similarly for other values denoting the Fourier transform by the same letter without hat. One can derive the analogue of \eqref{eq7} for the Fourier transform of the input fluctuation operators:
\begin{align}
\label{comm11}
\left[ a(\Omega),  a^\dag(\Omega')\right] &= 2\pi\,\delta(\Omega -\Omega'),\\
\label{corr11}
\left\langle a(\Omega)  a^\dag(\Omega')\right\rangle &= 2\pi\, \delta(\Omega -\Omega').
\end{align}

We assume that in \eqref{a1} the expected values exceed the
fluctuation parts of the operators. So we make use of the method
of successive approximation to derive a set of equations
describing the system. We select $A_0 = A_0^*$ and find the
following in zero order approximation:
\begin{align}
A_1=A, \quad A_0=\frac{2}{\sqrt{\gamma_0}}A.
\end{align}

One can find equations for the fluctuation part of the field and
the displacement of the test mass in the first order
approximation. in the spectral representation they have the
following form:
\begin{subequations}
    \label{aaapy}
\begin{align}
\label{eq10}
a_{1a}=&\frac{\frac{\gamma_0}{2}+i\Omega}{\frac{\gamma_0}{2}-i\Omega}a_a-\frac{i\Omega\eta A}{\frac{\gamma_0}{2}-i\Omega}\sqrt{2}y_{\Omega},\\
\label{eq11}
a_{1\phi}=&\frac{\frac{\gamma_0}{2}+i\Omega}{\frac{\gamma_0}{2}-i\Omega}a_{\phi}-\frac{2\omega_0\xi A}{\frac{\gamma_0}{2}-i\Omega}\sqrt{2}y_{\Omega},\\
\label{a0a}
a_{0a}=&\frac{\sqrt{\gamma_0}}{\frac{\gamma_0}{2}-i\Omega}a_{a}-\frac{\sqrt{\gamma_0}\eta A}{\sqrt{2}(\frac{\gamma_0}{2}-i\Omega)}y_{\Omega},\\
\label{a0p}
 a_{0\phi}=&\frac{\sqrt{\gamma_0}}{\frac{\gamma_0}{2}-i\Omega}a_{\phi}-\frac{2\omega_0\xi A}{\sqrt{\gamma_0}(\frac{\gamma_0}{2}-i\Omega)}\sqrt{2}y_{\Omega},\\
\label{yo}
y_{\Omega}=&-\frac{F_{\Omega}}{m\Omega^2}+\frac{2\sqrt{2}\hbar\omega_0\xi A}{\sqrt{\gamma_0}m\Omega^2}a_{0a}+\\&+\frac{\sqrt{\gamma_0}\eta\hbar A}{\sqrt{2}m\Omega^2}\left(\frac{2}{\sqrt{\gamma_0}}a_{\phi}-a_{0\phi} \right).
\end{align}
\end{subequations}
Here $y_{\Omega}$ and $F_{\Omega}$ are Fourier transform of the displacement $y$ and signal $F_s$ respectively, $a_a$ and $a_{\phi}$ are amplitude and phase quadratures that we define as follows
\begin{align}
\label{quad}
a_a &= \frac{a+ a_-^\dag}{\sqrt 2},\quad a_{\phi} = \frac{a - a_-^\dag}{i\sqrt 2},\quad a_-\equiv a(-\Omega).
\end{align}

From the equations \eqref{aaapy} we see that amplitude quadrature of the output field provides information about the speed of the probe mass $-i\Omega y_{\Omega}$, which corresponds to the dissipative coupling. In contrast, phase quadrature provides information about the displacement of the probe mass, which is typical for the dispersive coupling.

Let us substitute \eqref{a0a} and \eqref{a0p} into the equation for the spectrum of the displacement $y$ \eqref{yo}:
\begin{subequations}
    \label{eq12}
\begin{align}
&\left(K-m\Omega^2\right) y_{\Omega}=F_{\Omega}+F_{fl},\\
\label{Flf}
&F_{fl}=-\frac{2\sqrt{2}\hbar\omega_0\xi A}{\frac{\gamma_0}{2}-i\Omega}a_a-i\frac{\sqrt{2}\hbar\Omega\eta A}{\frac{\gamma_0}{2}-i\Omega}a_{\phi},\\
\label{KOmega}
&K(\Omega)=-\frac{4\hbar\omega_0\xi\eta A^2}{\frac{\gamma_0}{2}-i\Omega}\simeq \kappa-i\Omega\delta,\\
\label{rigidity}
&\kappa=-\frac{8\hbar\omega_0\xi\eta A^2}{\gamma_0},\quad \delta=\frac{16\hbar\omega_0\xi\eta A^2}{\gamma_0^2}.
\end{align}
\end{subequations}
Here $F_{fl}$ is the fluctuation back action force, $K(\Omega)$ is
the optical rigidity which is associated with both dissipative and
dispersive coupling ($K \sim \xi\eta$). Note that this rigidity
appears at resonance pump. Recall, in cases of pure dispersive
\cite{06PRAcckovwm} or pure dissipative \cite{19a1JPhyBNaVy}
coupling  optical rigidity is possible only in {\em detuned} pump.

We expand rigidity \eqref{KOmega} into the Taylor series over
$i\Omega$ keeping only two first terms (below we assume
$\gamma_0\gg\Omega$ \eqref{smallx}). This optical rigidity is
unstable. If $\kappa$ is positive then the mechanical viscosity
$\delta$ introduced is negative and vice versa. The rigidity
$\kappa$ is positive when $\xi\eta<0$. When $\kappa$ is positive
the probe mass effectively acts as a harmonic oscillator which is
affected by the signal $F_s$ and the fluctuation back action
force.

\begin{figure}
    \includegraphics[width=0.48\textwidth]{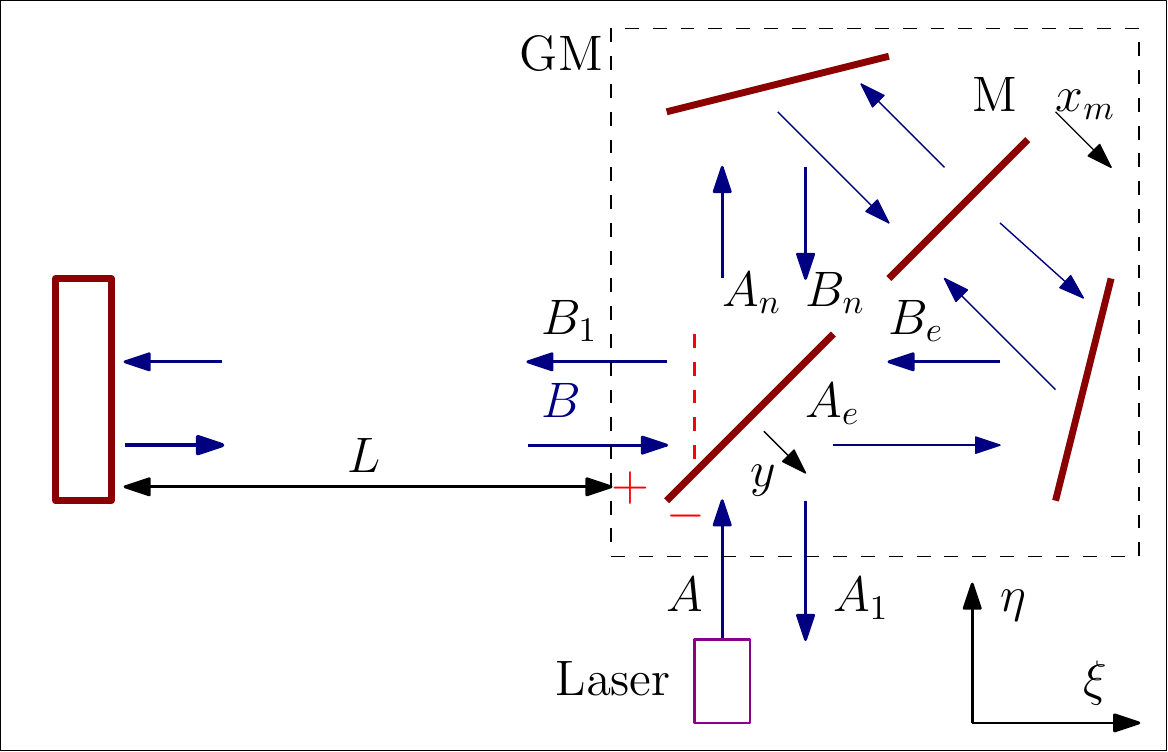}
    \caption{ Michelson-Sagnac interferometer as a generalized mirror (GM)  of FP cavity. Combined coupling takes place when beam splitter is movable but fixed mirror $M$ ($x_m$ --- const).}\label{Cavity}
\end{figure}

\section{Examples of realizations of combined coupling}
\label{exemple} For a realization of the combination of
dissipative and dispersive couplings we use the Michelson-Sagnac
interferometer (MSI) as one of the mirrors in the FP cavity (see
Fig.~\ref{Cavity}). The MSI consists of the 50/50 beam splitter
(BS) and three completely reflecting mirrors. This interferometer
can be considered as a generalized mirror having amplitude
transmittance $T$ and reflectivity $R$ depending on displacements
$z_m$ of the mirror $M$  and $y_{BS}$ of BS. So input-output
relations have the form (see notation on Fig.~\ref{Cavity}):
\begin{subequations}
    \label{MSIA1}
    \begin{align}
        \label{MSIA1a}
        \mathcal A_1 & =  \mathcal B T  + e^{-i k\sqrt 2y} \mathcal A R, \\
        \label{MSIA1b}
        \mathcal B_1 &=  \mathcal A  T - e^{i k\sqrt 2 y}\mathcal B R,\\
        \label{MSIA1cc}
        R &=   \cos k \big(2z_m+\sqrt 2 y\big), \\
        T & =  \sin k \big(2z_m + \sqrt 2 y\big),
    \end{align}
\end{subequations}
where $\mathcal A$ and $\mathcal B$ are amplitudes coherent monochromatic fields, $k=\frac{\omega}{c}$ is wave vector.

A detailed analysis of MSI is given in \cite{XuerebPRL2011,
TarabrinPRA2013, SawadskyPRL2015, 16a1PRAVyMa, 19a1JPhyBNaVy,
20PRAKaVy}. We assume that the waves passing through BS acquire a
phase shift equal to $\frac{\pi}{2}$, and the phases of the waves
reflected from BS are determined only by the displacement of the
BS itself.  We also assume that the spectral frequencies $\Omega$,
characterizing the displacements of BS and the moving mirror $M$
are small enough: $\Omega t_{in}\ll 1$, where $t_{in}$ is round
trip time of light between BS and mirror $M$. This means that the
circulating fields change phase almost instantly at small
displacements of BS and $M$. Amplitude transmittance $T$ and
reflectivity $R$ of the GM depend only on positions $z_m$, $y$.

Below we can designate displacement as
\begin{align}
    \label{xy}
    y\Rightarrow y_0 + y, \quad z_m\Rightarrow z_0+z,
\end{align}
where $z_0,\ y_0$ are the mean constants (to be chosen) and $z ,\ y$ are small variables.

Below we  consider only a situation of movable beam splitter and
the mirror $M$ fixed (that is $z=0$). Then we can expand $R,\ T$
\eqref{MSIA1} into a series
\begin{subequations}
    \label{expRT}
    \begin{align}
        \label{Rexp}
        R &  \simeq R_0 - T_0\, k \sqrt 2\, y \, , \\
        \label{Texp}
        T &  \simeq T_0 + R_0\, k \sqrt 2\, y\, ,\\
        R_0 &= \cos k  \big(2z_0 +\sqrt 2\, y_0\big),\quad T_0=  \sin k  \big(2z_0 +\sqrt{2}y_0\big),\nonumber
    \end{align}
\end{subequations}
For simplicity we put $y_0=0$ below, then only $z_0$ defines $T_0,\ R_0$.

The MSI is a part of the FP cavity. The field $B_1$ propagates to the end mirror (we consider it is completely reflecting), reflects and comes back to the beam splitter. In the stationary mode of operation fields $B$ and $B_1$ have the following coupling
\begin{align}
    \label{B}
    \mathcal B=\mathcal B_1e^{i2kL},
\end{align}
where $L$ is the distance between the beam splitter and the end
mirror.

Let's substitute \eqref{B} in equations \eqref{MSIA1}
\begin{subequations}
    \label{A1B1}
    \begin{align}
        \mathcal A_1 & = \mathcal B_1e^{i2kL} T  + e^{-i k\sqrt 2y} \mathcal A R,\\
        \label{B1}
        \mathcal B_1 & = \mathcal A  T - e^{i k\sqrt 2 y}\mathcal B_1e^{i2kL} R
    \end{align}
\end{subequations}

The internal field's power $I_0$ is given by the ratio
\begin{align}
    \label{I0}
 I_{0} = \frac{T^2I}{1+R^2+2R\cos2k\left(L+\frac{y}{\sqrt{2}} \right) }.
\end{align}
Where $I$ is the input field power.

This power achieve a maximum on $e^{2ik(L+y/\sqrt{2})}=-1$ (here and below we assume that $R>0$). We can find the resonant frequency from this equation
\begin{align}
    \label{wr}
w_r=\frac{\omega_0}{1+\frac{y}{\sqrt{2}L}}\simeq \omega_0\left(1-\frac{y}{\sqrt{2}L} \right).
\end{align}
Here $\omega_0$ is the resonant frequency by $y=0$.

Now let's find the half bandwidths of the cavity $\gamma_0$. Let
us assume that the wave vector $k$ in equation \eqref{I0} is equal
to $(\omega_r+\gamma)/c$ ($\omega_r\gg\gamma$) so that
$I_0=I_{0\text{max}}/2$. Then we get the following relation for
$\gamma$
\begin{align}
    \label{g}
    \gamma=\frac{2\left(1-R\right)}{\tau\sqrt{R}}\simeq\frac{T^2}{\tau}\simeq\frac{T_0^2}{\tau}\left(1+\frac{2\sqrt{2}k_0R_0}{T_0} \right).
\end{align}
Here we account that $T_0 \ll 1$.

Let's compare equations \eqref{wr} and \eqref{g} with
\eqref{Hamilt}. We see that coefficients of dispersive ($\xi$) and
dissipative ($\eta$) coupling for the system described above have
the following form
\begin{align}
    \label{xieta}
    \gamma_0=\frac{T_0^2}{\tau}, \quad \xi=-\frac{1}{\sqrt{2}L}, \quad \eta=\frac{2\sqrt{2}k_0}{T_0}.
\end{align}

In Table \ref{t1} we list parameters of the system described
above, which can be used in a laboratory experiment.

Another example of realizing combined coupling is given in
\cite{SawadskyPRL2015, SawadskyArxiv2015}. The authors described
an optical-mechanical system similar to the one presented above,
but the test mass was a partially transmitting mirror M and beam
splitter was immobile (see Fig.~\ref{Cavity}). For this case the
input-output relations can be written as follows
 \begin{subequations}
    \begin{align}
            \mathcal A_1 & =  i\mathcal B \mathbb{T}  + \mathcal A \mathbb{R}_{\triangleright}, \\
        \mathcal B_1 &=  i\mathcal A \mathbb{T} +\mathcal B \mathbb{R}_{\triangleleft},\\
        \mathbb{T} & =  e^{ikl_+}r_{M}\sin 2k\delta l,\\
        \mathbb{R}_{\triangleright} &= e^{ikl_+}\left(r_M\cos 2k\delta l - it_M \right), \\
        \mathbb{R}_{\triangleleft} &= e^{ikl_+}\left(r_M\cos 2k\delta l + it_M \right).
    \end{align}
 \end{subequations}
Here $r_M$ and $t_M$ are the  amplitude reflectivity and transmittance of mirror, $l_+$ and $\delta l$ are the sum and difference of the lengths of the MSI arms, respectively. We consider that $\delta l = z_0+z$, where $z_0$ is a constant, and $z$ is a small displacement, $kz \ll 1$.

Now we can find the coefficients of dispersive $\xi_1$ and dissipative $\eta_1$ coupling for this optomechanical system by conducting the analysis presented above
\begin{align}
 \label{xi1eta1}
    \gamma_1=\frac{r_M^2T_1^2}{\tau}, \quad \xi_1=-\frac{T_1t_Mr_M}{L}, \quad \eta_1=\frac{4k_0}{T_1}.
\end{align}
Here we assume that $|T_1|=|\sin 2kz_0| \ll 1$ and find resonant frequency $\omega_0$ from $e^{ik\left( 2L+l_+\right)  + i\phi_0}=1$, where $\phi_0 = \arctan\left(  t_M/r_M\right) $.

These examples show that we can choose coupling coefficients $\xi,\ \eta$ within some bounds.

\begin{table}
    \caption{\label{t1}Parameters of the optomechanical system}
    \begin{center}
        \begin{tabular}{cc}
            \hline
            Parameter&Value\\
            \hline
            Medium amplitude transmittance\\ of MSI $T_0$&0.01\\
            Probe mass $m$& 50 g\\
            Pump frequency $
            \omega_0/2\pi$& 300 THz\\
            Pump power $I_0$& 42 mW\\
            Cavity length $L$& 1 m\\
            Cavity half bandwidth $\gamma_0$&15000 $\text{s}^{-1}$\\
            Coefficient of the dissipative coupling $\eta$&$1.78\times10^9$ $\text{m}^{-1}$\\
            Coefficient of the dispersive coupling $\xi$&$-0,71$ $\text{m}^{-1}$\\
            \hline
        \end{tabular}
    \end{center}
\end{table}

\section{Detecting signal force}

We consider an optomechanical system with a combination of both couplings $\xi,\ \eta$ as a signal force detector, assuming that  $\xi,\ \eta$ can be varied {\em arbitrary}. Let's find the sensitivity of this measurement.

We assume that quadratures of the output field \eqref{eq10},
\eqref{eq11} are processed optimally for this purpose. Let's
substitute \eqref{eq12} in equations of quadratures \eqref{eq10}
and \eqref{eq11}:
\begin{subequations}
    \label{eq13}
    \begin{align}
        \label{aa}
    &a_{1a}=\frac{(x_0^2-x^2)a_a-P_mx^2a_{\phi}-ix^2\sqrt{2P_m}f_s}{(x_0^2-x^2)-ix\delta_m},\\
    \label{ap}
    &a_{1\phi}=\frac{(x_0^2-x^2)a_{\phi}+Q_mD^2a_{a}-xD\sqrt{2Q_m}f_s}{(x_0^2-x^2)-ix\delta_m},\\
    \label{pq}
    &P_m=\frac{8\hbar\eta^2A^2}{m\gamma_0^2}, \quad Q_m=\frac{32\hbar\xi^2A^2}{m\gamma_0^2},\\
    \label{omega}
    &\delta_m=\frac{\delta}{m\gamma_0}= D\sqrt{P_mQ_m},  \quad x=\frac{\Omega}{\gamma_0},\quad x_0=\frac{\Omega_0}{\gamma_0}\nonumber\\
    &\Omega_0=\sqrt\frac{ \kappa}{m},,\quad
    f_s=\frac{F_s}{\sqrt{2\hbar m\Omega^2}}, \quad D=\frac{\omega_0}{\gamma_0}.
    \end{align}
\end{subequations}
Here $f_s$ is the signal force normalized to SQL, $\kappa$ and
$\delta$ are given by the equations \eqref{rigidity}, $\Omega_0$
is the resonant frequency which appears due to the optical
rigidity \eqref{rigidity}, $D$ is the quality factor of the
optical cavity. Here and below we assume that
$\xi\eta=-|\xi\eta|$ and
\begin{align}
 \label{smallx}
 \Omega,\, \Omega_0 \ll \gamma_0, \quad \text{or}\quad  x,\, x_0\ll 1
\end{align}

It follows from equations \eqref{eq13} that both quadratures are suitable for detecting the signal force. We use the homodyne detection for the measurement of quadratures, allowing to measure a quadrature
\begin{align}
 \label{atheta}
 a_{1\theta}=a_{1a}\cos\theta+a_{1\phi}\sin\theta,
\end{align}
where $\theta$ is a homodyne angle.  Let input field is in
coherent state. This means that single-sided power spectral
densities (PSD) of quadratures $a_a$ and $a_\phi$ equal
$S_a(\Omega) =S_\phi(\Omega)=1$ \cite{02a1KiLeMaThVyPRD}. Then
noise PSD recalculated to $f_s$ can be easily derived from
\eqref{eq13}:
\begin{subequations}
    \label{Sf}
    \begin{align}
    S_{f}&=S_{a1}+S_{\phi 1},\\
    S_{a1}&=\frac{\left( x_0^2-x^2+Q_mD^2\tan\theta\right)^2 }{2x^2\left(x^2P_m+D^2Q_m\tan^2\theta \right) },\\
    S_{\phi1}&=\frac{\left((x_0^2-x^2)\tan\theta-P_mx^2 \right)^2}{2x^2\left(x^2P_m+D^2Q_m\tan^2\theta \right)}.
    \end{align}
\end{subequations}
Here SQL sensitivity corresponds to  $S_{f}=1$. When $\theta=0$ we
measure the amplitude quadrature, and when $\theta=\frac{\pi}{2}$
we measure the phase quadrature.

Let's fix $x=x_c$ and find $\tan\theta$ at which equation
\eqref{Sf} takes an extreme value. There are two $\tan\theta$ and
they have the following form
\begin{align}
    \tan\theta_1=-\frac{x_0^2-x_c^2}{Q_m D^2},\quad \tan\theta_2=\frac{P_mx_c^2}{x_0^2-x_c^2}.
\end{align}

Choosing these homodyne angles we completely cancel the noise
determined by one of the quadratures, namely, $S_{a1}^{ x=x_c}=0$
at $\theta=\theta_1$ and $S_{\phi1}^{ x=x_c}=0$ at
$\theta=\theta_2$.

Let's consider a special case $x_c=x_0$. Then $\tan\theta_1=0$ and
$\tan\theta_2=\pm\infty$. This means that we measure the
quadrature of the amplitude or phase. In these measurements PSDs
 have the following form

\begin{subequations}
\begin{align}
    \label{Sfa}
    &S_{f}|^{\theta=0}=\frac{1}{2}\left[P_m+\frac{\left(\left(\frac{x_0}{x} \right)^2-1  \right)^2 }{P_m} \right],\\
    \label{Sfp}
    &S_f|^{\theta=\frac{\pi}{2}}=\frac{1}{2x^2}\left[Q_mD^2+\frac{(x_0^2-x^2)^2}{Q_mD^2} \right].
\end{align}
\end{subequations}

In resonance case $\Omega=\Omega_0$
\begin{subequations}
    \label{minSfa}
\begin{align}
    \label{eq15}
S_{f}(\Omega_0)|^{\theta=0} &=\frac{P_m}{2}=x_0^2g,\quad g=\sqrt\frac{P_m}{Q_mD^2}\\
    \label{eq16}
S_{f}(\Omega_0)|^{\theta=\pi/2} &=\frac{D^2Q_m}{2x_0^2}=\frac{1}{g}.
\end{align}
\end{subequations}
Here we rewrite the PSD using equations \eqref{pq} and \eqref{omega}, $g$ is a ratio  between coefficients of optomechanical couplings.

The relations \eqref{minSfa} show that at $g\ll 1/\sqrt{x_0}$ we
have inequality  $S_{f}(\Omega_0)|^{\theta=0}\ll
S_{f}(\Omega_0)|^{\theta=\pi/2}$ and in order to surpass SQL near
the resonant frequency we have to detect amplitude quadrature
(i.e., $\theta=0$). In the opposite case $g\gg 1/\sqrt{x_0}$ we
have inverse inequality  $S_{f}(\Omega_0)|^{\theta=0}\ll
S_{f}(\Omega_0)|^{\theta=\pi/2}$ and to surpass SQL one has to
detect the phase quadrature (i.e., $\theta=\pi/2$).

 \begin{figure}
    \includegraphics[width=0.5\textwidth]{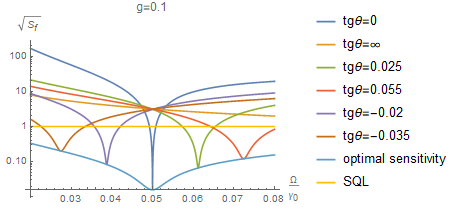}
    \includegraphics[width=0.5\textwidth]{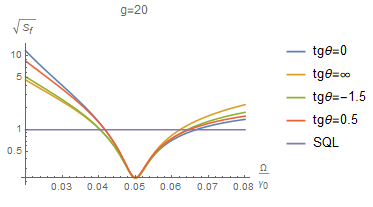}
    \includegraphics[width=0.5\textwidth]{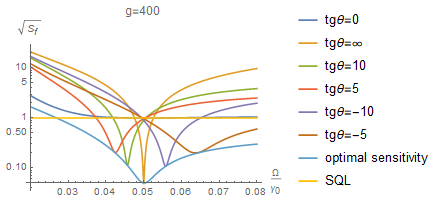}
    \caption{Graphs of amplitude spectral densities $\sqrt{S_{f}(\Omega)}$ plotted for the homodyne detection having different homodyne angles and ratio $g$ and fixed dimensionless frequency $x_0=0.05$.  The upper graphs are obtained for $g=0.1$ The middle graphs correspond to condition \eqref{S=S} $g=1/x_0=20$. The graphs at the bottom are plotted for the case $g=400\gg 1/x_0$. Also upper and bottom graphs show the optimal PSD by the optimal frequency-dependent homodyne angle (see in Appendix \ref{squeez}).}\label{ASD1}
\end{figure}

Recall, $S_{f}(\Omega_0)|^{\theta=0}$ and
$S_{f}(\Omega_0)|^{\theta=\pi/2}$ are extremes of one function.
The maximum (minimim) depends on ratio $g$. However, at $g=1/x_0$
they became equal to each other (the maximum and the minimum
coincide):
\begin{align}
 \label{S=S}
 S_{f}(\Omega_0)|^{\theta=0}&=S_{f}(\Omega_0)|^{\theta=\pi/2}= x_0, \quad \text{at }
    g=\frac{1}{x_0}
\end{align}
Thus, when $g=1/x_0$ we get the same sensitivity near the resonant
frequency for any quadrature detection. Usually $x_0\ll 1$ (the
case of non-resolved sideband), hence, SQL can be surpassed.

The minimum PSD is achieved at the resonant frequency and it is
defined by equations \eqref{minSfa} . This minimum PSD is realized
inside a narrow bandwidth $\Gamma$:
\begin{align}
    \label{eq14}
    \frac{\Gamma}{\Omega_0}  \simeq S_{f}^\text{min}.
\end{align}
Here $\Gamma$ is defined as $S_{f}(\Omega_0 \pm \Gamma/2) \simeq 2 S_{f}^\text{min}$. The relation \eqref{eq14} corresponds to the known Cramer-Rao bound \cite{Mizuno, Mizuno93, Miao2017}.

In Fig.~\ref{ASD1} we depict graphs of amplitude spectral
densities $\sqrt{S_{f}(\Omega)}$ of noise recalculated to $f_s$ by
the homodyne detection with different homodyne angles and ratio
$g$ and fixed dimensionless frequency $x_0=0.05$ and pump power
(so $|\xi\eta|=1$). The main parameters (test mass, optical power)
are taken from Table~\ref{t1} with varying coupling $\xi,\ \eta$. The plots are given for preliminary chosen fixed homodyne angle. For methodical purpose on upper and bottom graphs we also show the optimal PSD by the optimal frequency-dependent homodyne angle (see in Appendix \ref{squeez}).

The upper graphs in Fig.~\ref{ASD1} are obtained for $g=0.1$.
Varying homodyne angle one can surpass SQL at frequencies close to
mechanical resonance. At mechanical resonance the sensitivity
attains the minimum but in narrow bandwidth. SQL can be surpassed
when the frequency differs from the resonance one by about 50\%,
but sensitivity will by slightly drop and the bandwidth will be
wider if compared to the case of mechanical resonance.

The middle graphs in Fig.~\ref{ASD1} correspond to condition
\eqref{S=S} $g=1/x_0=20$. SQL can be surpassed within a relatively
wider bandwidth (about 50\% of center frequency). Variation of
homodyne angle practically does not influence on sensitivity.

And the bottom graphs in Fig.~\ref{ASD1} describe the case
$g=400\gg 1/x_0$. Variation of homodyne angle allows surpassing
SQL inside the bandwidths close to the mechanical resonance. It is
similar to the case shown in upper graphs.

Above we assumed that  $\xi,\ \eta$ can be varied {\em arbitrary}.
But usually these coefficients are constant in certain systems.
For example, the Fabry-Perot cavity with the MSI with the movable
BS has fixed $\xi$ and $\eta$ (see \eqref{xieta}), and the ratio
$g=T_0 \ll 1$.

If only the  partially transmitting mirror M is movable in the
MSI, then coefficients \eqref{xi1eta1} are constant too, and
$g=r_M/t_M$. In this case we can get an optomechanical system with
a large coefficient $g$ if $r_M \gg t_M$.

\section{Ponderomotive  squeezing}

We would like to pay attention to the fact that combined
optomechanical coupling can also be used to produce a pondermotive
squeezed light. In turn, varying ratio $g$ \eqref{eq15} between
dispersive and dissipative coupling provides for a possibility to
control output  squeezing. The output quadrature \eqref{atheta}
measured by homodyne detector can be derived from \eqref{eq13}:
\begin{subequations}
 \label{atheta2}
 \begin{align}
 a_{1\theta} &= \frac{ (x_0^2-x^2)\cos \theta+ \frac{2x_0^2}{g}\sin\theta }{\left( x_0^2-x^2\right) - 2ixx_0^2 }\, a_a+\\
    &\quad +\frac{\left(x_0^2-x^2\right)\sin\theta-2x_0^2x^2g \cos\theta}{\left( x_0^2-x^2\right) -2i x x_0^2 }\, a_\phi
\end{align}
\end{subequations}
In the case of the mechanical resonance $\Omega=\Omega_0$ this
equation can be written as follows:
\begin{align}
a_{\theta}|_{x=x_0}=\frac{i}{gx_0}\sin\theta\, a_{a}-ix_0g\cos\theta\, a_{\phi};
\end{align}

Obviously, in order to measure squeezing for small $g\ll 1/x_0$
one has to choose $\sin \theta=0$ and to measure amplitude output
quadrature. In contrast, to measure squeezing for large  $g\gg
1/x_0$ one has to choose $\cos \theta=0$ and to measure the phase
output quadrature (see \eqref{PSD} in Appendix \ref{squeez}):
\begin{subequations}
\begin{align}
S_{a}(\Omega_0)&=x_0^2g^2, \quad (\sin \theta=0)\\
S_{\phi}(\Omega_0)&=\frac{1}{g^2x_0^2}, \quad (\cos \theta =0),
\end{align}
\end{subequations}
where $S_{a}$ and $S_{\phi}$ are single-sided PSD of output
amplitude and phase quadratures, respectively, and we assume that
condition \eqref{smallx} is valid. Such squeezing near the resonant
frequency $\Omega_0$ is not observed in the case of dispersion
coupling.

\begin{figure}
    \includegraphics[width=0.5\textwidth]{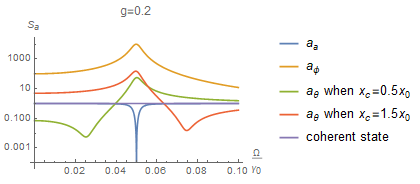}
    \includegraphics[width=0.5\textwidth]{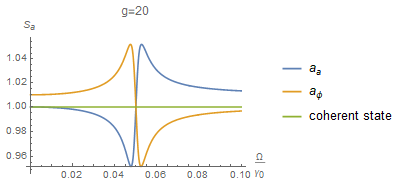}
    \includegraphics[width=0.5\textwidth]{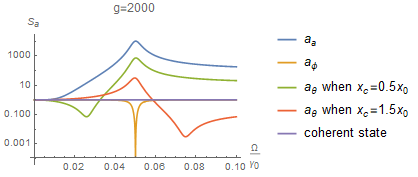}
    \caption{Graphs of single-sided PSD of different quadratures by different parameter $g$ and the fixed resonant frequency $x_0=0.05$. The upper graphs are constructed for $g=0.2\ll 1/x_0$. The middle graphs are constructed for  $g=1/x_0=20$, here we get an almost coherent state of the electromagnetic field (pay attention to scale on vertical axis). And bottom graphs relate to case $g=2000\gg 1/x_0$.}\label{PMsq}
\end{figure}

In case \eqref{S=S} $g=1/x_0$ the output light is practically
coherent and it is required to pay attention to scale on vertical
axis of middle plots on Fig.~\ref{PMsq}.

Specifically, in the case of low frequencies $\Omega\ll\Omega_0$
($x\ll x_0$) we get frequency-independent squeezing:
\begin{align}
 \label{combD1}
  \Omega\ll \Omega_0\quad g\ll 1 :\ S_{\theta}^{\text{comb}}\simeq\frac{g^2}{4},
\end{align}
here $S_\theta^{\text{comb}}$ is a single-sided PSD of the output
quadrature \eqref{atheta} at optimal homodyne angle in case of
combined coupling (see details in Appendix \ref{squeez}).

It is similar to what we obtained in the case of the dispersive
coupling with a non-resonant pump ($\Delta= \omega_p -\omega_0$ is
detuning) \cite{06PRAcckovwm}:
 \begin{align}
 \label{combD2}
  \Omega\ll \Omega_0, \quad \frac{\Delta}{\gamma_0}\ll 1:\
  S_\theta^{\text{disper}}\simeq\frac{\Delta^2}{\gamma_0^2},
\end{align}
where  $S_\theta^{\text{disper}}$ are single-sided PSD of output
quadrature \eqref{atheta} at optimal homodyne angle  for
dispersive coupling.  The PSD (\ref{combD1}, \ref{combD2}) are
equal to each other when $\frac{g}{2}=\frac{\Delta}{\gamma_0}$ and
practically do not depend on the frequency.

In general case we can adjust the maximum squeezing at preliminary
chosen dimensionless frequency $x_c$ (near $x_0$) by varying
homodyne angle $\theta$ at a fixed ratio $g$ (see details of
calculations in Appendix \ref{squeez}). If compared with
dispersive case the main advantage of combined coupling deals with
possibility to vary degree of squeezing and its bandwidth choosing
$x_c$ (i.e., the homodyne angle).

Shown in Fig.~\ref{PMsq} are the plots of single-sided PSD
\eqref{PSDxc} obtained for the output light for normalized
mechanical frequency $x_0=0.05$ when $x_c=0.5 x_0$ or $x_c=1.5
x_0$ and different ratio $g$.  For  $g=1/x_0$ (the middle plot in
Fig.~\ref{PMsq}) the output state is practically coherent at
$\Omega=\Omega_0$ but out of resonance in case of squeezing. But
the farther $g$ is from $1/x_0$, the stronger the squeezing
becomes near $x_0$. For small $g \ll 1/x_0$ we get the amplitude
quadrature squeezed near the resonance, and for large $g \gg
1/x_0$ the phase quadrature. It is possible to get less strong
squeezing but in a wider frequency band at other frequencies. This
can be done measuring a quadrature $a_{1\theta}$ other than the
amplitude and phase.

\section{Discussion and conclusion}

We analysed the optomechanical system, featuring combination of
dispersive and dissipative coupling, and showed that the main
properties of combined coupling deal with the optical rigidity
\eqref{eq12}, which appears as a result of {\em both} kinds of
coupling. At the same time, this rigidity manifests in systems
having combined coupling on the resonant pump, which is not
typical for pure dispersive \cite{06PRAcckovwm} or dissipative
\cite{16a1PRAVyMa} coupling types.

For realizing the combination of dissipative and dispersive
coupling we used the MSI as an input mirror in the FP cavity (see
Fig.~\ref{Cavity}). We considered two different modes of operation
of the MSI with a movable beam splitter and an immobile
completely reflecting mirror M and vice versa with a movable
partially transmitting mirror M and a fixed beam splitter. The
coefficients of the dispersive $\xi$ and dissipative $\eta$
coupling for these schemes were obtained. In further analysis, we
assumed that $\xi$ and $\eta$ can be varied arbitrary.

We considered an optomechanical system with a combination of both couplings as a signal force detector. Homodyne detection of an output field can have a sensitivity better than SQL near the resonant frequency $\Omega_0$, which is defined by the optical rigidity \eqref{rigidity}. The analysis shows that it is most effective to measure the amplitude or phase quadrature (the choice of the quadrature depends on the ratio $g$ as compared with $1/x_0$ \eqref{S=S}). If $g<1/x_0$, it is better to measure the amplitude quadrature, and if $g>1/x_0$ --- the phase quadrature. At $g=1/x_0$ we get the same sensitivity  near the resonant frequency for measuring {\em any} quadrature. In this case the PSD recalculated to force SQL \eqref{S=S} is smaller than unity (i.e. SQL can be surpassed) if $x_0=\Omega_0/\gamma_0\ll 1$.

The physical reason is related to the correlation between the measurement noise and the fluctuation back action. This correlation occurs due to the combination of both optomechanical couplings. Indeed, the fluctuation back action force \eqref{Flf} depends on the phase quadrature (dissipative coupling) and on the amplitude quadrature (dispersive coupling) and the measurement noise is determined by the amplitude or phase quadrature (first terms in Eq. \ref{eq10} and \ref{eq11}). Part of the total noise is completely compensated at the resonant frequency (see Eq. \ref{eq13}). The remaining noise recalculated to $f_s$ is proportional to $\sqrt{P_m}$ or $\sqrt{D^2Q_m}/x_0$, depending on which quadrature we measure. SQL is surpassed if this noise is small.

We would like to point out that variation of ratio $g$ between dispersive and dissipative coupling  and choice of homodyne angle provide possibility to control output poderomotive squeezing. Varying homodyne angle we can obtain constant squeezing at frequencies much smaller than resonant frequency $\Omega_0$, or
large squeezing in finite bandwidth (the larger squeezing, the more narrow bandwidth) near the resonant frequency. The ponderomotive squeezing induced by combined coupling has a wider range of varying  squeezing parameters as compared with ponderomotive squeezing caused by dispersive coupling. 

The combined coupling looks promising to be used in gravitational wave antennas for creation of frequency dependent squeezing with controllable parameters. The main obstacle is thermal mechanical noise. It is the subject of our future research.

 \acknowledgments
The authors are pleased to Haixing Miao for fruitful discussion and advises. 
They are grateful for support provided by the Russian Foundation for Basic Research (Grant No. 19-29-11003), the Interdisciplinary Scientific and Educational School of M.V. Lomonosov Moscow State University ``Fundamental and Applied Space Research''  and for TAPIR GIFT MSU Support from the California Institute of Technology. This document has LIGO number P2100446-v1.

\appendix
\section{The analysis of the pondermotive squeezing.}\label{squeez}
From \eqref{atheta2} one can calculate single-sided PSD assuming coherent input light
\begin{subequations}
 \label{PSD}
 \begin{align}
  S_\theta &= \frac{\left(x_0^2-x^2\right)^2+ 4x_0^4\left(\frac{\sin^2\theta}{g^2} +x^4 g^2\cos^2\theta\right)
    }{ \left(x_0^2-x^2\right)^2+4x^2x_0^4}+\\
    &\quad + \frac{4x_0^2 \left(x_0^2-x^2\right)\sin\theta\cos\theta\left(\frac{1}{g} -x^2g\right)}{
     \left(x_0^2-x^2\right)^2+4x^2x_0^4}=\\
   &= \frac{W + U \cos 2\theta + V\sin 2\theta  }{
     \left(x_0^2-x^2\right)^2+4x^2x_0^4}\\
  W &= \left(x_0^2-x^2\right)^2 + \frac{2x_0^4}{g^2} +2x^4x_0^4 g^2 ,\\
  U &=  2x_0^2\left[x^4x_0^2 g^2-\frac{x_0^2}{g^2} \right],\\
  V &= 2x_0^2\left(x_0^2-x^2\right) \left[\frac{1}{g} -x^2g\right]
   \end{align}
  \end{subequations}

 The minimum of $S_\theta$ at $x=x_c$ takes place at homodyne angle $\theta$ defined as
\begin{subequations}
 \label{PSD3}
 \begin{align}
  \cos2\theta &= -\frac{U_c}{\sqrt{U_c^2+V_c^2}},\quad  \sin2\theta = -\frac{V_c}{\sqrt{U_c^2+V_c^2}},\\
  U_c &= U|_{x=x_c},\quad V_c = V|_{x=x_c}.
 \end{align}
 and it is equal to
 \begin{align}
  \label{PSDxc}
  S_\theta^{x_c}(x) &=  \frac{W - \frac{UU_c +VV_c}{\sqrt{U_c^2+V_c^2}}  }{
     \left(x_0^2-x^2\right)^2+4x^2x_0^4}
 \end{align}

\end{subequations}

 In particular case $x \ll x_0$ we have
\begin{subequations}
    \label{PSD2}
    \begin{align}
    W &\simeq  x_0^4\left(1 + \frac{2}{g^2}\right),\quad
    U \simeq  -\frac{2x_0^4}{g^2} ,\quad
    V \simeq  \frac{2x_0^4}{g},\\
    S_\theta &^{x\ll x_0}\simeq 1 + \frac{2}{g^2} -\frac{2}{g^2}\cos2\theta + \frac{2}{g}\sin 2\theta\ge\\
    &   \ge \frac{\sqrt{1+g^2} -1}{\sqrt{1+g^2} + 1}
    = \frac{g^2}{4} -\frac{g^4}{8} +\dots
    \end{align}

\end{subequations}

\bibliographystyle{ieeetr}
\bibliography{OptoMech.bib}

\end{document}